\documentclass[11pt]{article}
\usepackage{amsmath,amssymb}
\usepackage{epsfig}
\begin{document}
\begin{center}
   \vskip 3em
{\LARGE {\bf Solution of the Helmholtz equation }  }
\vskip 1em
{\LARGE {\bf for spin-2 fields}}
  \vskip 3em
{\large   G. F. Torres del Castillo$^a$ \footnote{Email: gtorres@fcfm.buap.mx} \,  and
J. E. Rojas Marcial$^b$ \footnote{Email: efrojas@uv.mx}
 \\
{\small\it $^a$ Departamento de F\'\i sica Matem\'atica, Instituto de Ciencias,}\\
{\small \it Universidad Aut\'onoma de Puebla}\\
{\small\it  7200 Puebla, Pue., M\'exico}\\
{\small\it $^b$ Facultad de Ciencias F\'\i sico Matem\'aticas,} \\
{\small \it Universidad Aut\'onoma de Puebla} \\
{\small\it Apartado postal 1152, Puebla, Pue., M\'exico}
}
\end{center}
\vskip 2em
\begin{abstract}
The Helmholtz equation for symmetric, traceless, second-rank tensor fields
in three-di\-men\-sio\-nal flat space is solved in spherical and cylindrical  coordinates
by separation of variables making use of the corresponding spin-weighted harmonics.
It is shown that any symmetric, traceless, divergenceless second-rank tensor field
that satisfies the Helmholtz equation can be expressed in terms of two scalar potentials
that satisfy the Helmholtz equation. Two such expressions are given, which are
adapted to the spherical or cylindrical coordinates. The application to the linearized
Einstein theory is discussed.
\end{abstract}
\newpage
\setcounter{footnote}{0}

\section{Introduction}
The gravitational field in the linearized Einstein theory can be represented
by two symmetric, traceless, second-rank tensor fields which, assuming that the
fields vary harmonically in time, obey the Helmholtz equation outside the sources.
As in the case of other nonscalar equations, the solution of the Helmholtz equation
for second-rank tensor fields in noncartesian coordinates is a difficult problem
owing to the coupling of the fields components. However, when a nonscalar equation
is written in spherical or cylindrical coordinates, a considerable simplification
can be obtained by using spin-weighted quantities and the spin-weighted harmonics.

In this paper the Helmholtz equation for spin-2 fields ({\it i.e.,} symmetric,
traceless, second-rank tensor fields) is solved by separation of variables in spherical
and cylindrical coordinates making use of the spin-weighted harmonics. A similar
treatment for the vector (spin-1) Helmholtz equation is given in Refs. \cite{Torres1,
Torres2, Torres3}. In Sect. 2 the Helmholtz equation for spin-2 fields is solved in
spherical coordinates and it is shown that the divergenceless solutions of this equation
can be expressed in terms of two scalar (Debye) potentials that satisfy the
Helmholtz equation. The expressions for the divergenceless solutions in terms
of potentials obtained here are equivalent to those found in Refs. \cite{Campbell, Torres4}
for the multipoles with $j > 1$. In Sect. 3 the Helmholtz equation for
spin-2 fields is solved in cylindrical coordinates and an expression for the
divergenceless solutions in terms of potentials adapted to the cylindrical
coordinates is obtained. In Sect. 4 two alternative expressions for the solutions
of the linearized Einstein vacuum field equations in terms of scalar potentials
are given.

\section{Separation of variables in spherical coordinates}
\setcounter{equation}{0}

Let $\{ {\mbox{\bf e}}_1 , {\mbox{\bf e}}_2 ,{\mbox{\bf e}}_3 \}$ be an
orthonormal basis, a quantity $\eta$ has spin-weight $s$ if under
the transformation
\begin{equation}
{\mbox{\bf e}}_1 + i {\mbox{\bf e}}_2 \rightarrow e^{i \alpha}
\left( {\mbox{\bf e}}_1 + i {\mbox{\bf e}}_2 \right)
\end{equation}
it transforms according to
\begin{equation}
\eta  \rightarrow e^{i s \alpha} \eta.
\end{equation}
The five independent components of a symmetric, traceless tensor, $t_{ij}$,
can be combined to form the quantities
\begin{eqnarray}
t_{\pm 2} &\equiv & \frac{1}{2}\left( t_{11} - t_{22} \pm 2i t_{12} \right)
= \frac{1}{2}(t_{33} + 2t_{11} \pm 2i t_{12}),
\nonumber \\
 t_{\pm 1} &\equiv & \mp \frac{1}{2}\left( t_{13} \pm i t_{23} \right),
\label{eq:components} \\
t_0 &\equiv &  \frac{1}{2}t_{33}, \nonumber
\end{eqnarray}
so that $t_s$ has spin-weight $s$. If the components $t_{ij}$ are real, then
\begin{equation}
\bar{t_s} = (-1)^s t_{-s},
\label{eq:identity}
\end{equation}
where the bar denotes complex conjugation. Similarly, in the case of a
vector field  $F_i$, the combinations
\begin{equation}
F_{\pm 1} \equiv \pm \frac{1}{\sqrt{2}}\left( F_1 \pm iF_2 \right),
\quad \quad F_0 \equiv - \frac{1}{\sqrt{2}} F_3,
\label{eq:efes}
\end{equation}
have spin-weight $\pm 1$ and 0.

By choosing the basis  $\{ {\mbox{\bf e}}_1 , {\mbox{\bf e}}_2 ,{\mbox{\bf e}}_3 \}$
as the basis $\{ {\mbox{\bf e}}_\theta , {\mbox{\bf e}}_\phi ,{\mbox{\bf e}}_r \}$
induced by the spherical coordinates, one finds that the Helmholtz equation
for a symmetric, traceless tensor field $t$,
\begin{equation}
\nabla^2 t + k^2 t = 0,
\label{eq:helmholtz}
\end{equation}
written in terms of the spin-weighted components (\ref{eq:components})
amounts to
\begin{eqnarray}
\frac{1}{r^2} \partial_r (r^2 \partial_r t_{+2}) + \frac{1}{r^2} \eth {\bar{\eth}}
t_{+2} +  \frac{4}{r^2} \eth t_{+1} + k^2 t_{+2} &=&0 , \nonumber \\
\frac{1}{r^2} \partial_r (r^2 \partial_r t_{+1}) - \frac{4}{r^2} t_{+1}
+ \frac{1}{r^2} \eth \bar{\eth} t_{+1} -  \frac{1}{r^2} \bar{\eth} t_{+2}
+ \frac{3}{r^2} \eth t_0 + k^2 t_{+1} &=&0 , \nonumber \\
\frac{1}{r^2} \partial_r (r^2 \partial_r t_0) - \frac{6}{r^2} t_0
+ \frac{1}{r^2} \eth \bar{\eth} t_{0} +  \frac{2}{r^2}
\left( \eth t_{-1} - \bar{\eth}t_{+1} \right)
+  k^2 t_0 &=&0 ,\label{eq:splitted} \\
\frac{1}{r^2} \partial_r (r^2 \partial_r t_{-1}) - \frac{4}{r^2} t_{-1}
+ \frac{1}{r^2}  \bar{\eth} \eth t_{-1} + \frac{1}{r^2} \eth t_{-2}
- \frac{3}{r^2} \bar{\eth} t_0 + k^2 t_{-1} &=&0 ,
\nonumber \\
 \frac{1}{r^2} \partial_r (r^2 \partial_r t_{-2}) + \frac{1}{r^2} \bar{\eth}\eth
t_{-2} -  \frac{4}{r^2} \bar{\eth} t_{-1} + k^2 t_{-2} &=&0 ,\nonumber
\end{eqnarray}
where, acting on a quantity $\eta$ with spin-weight $s$ \cite{Newman,Torres5},
\begin{eqnarray}
\eth \eta &\equiv& - \sin^s \theta \left( \partial_\theta
+ \frac{i}{\sin \theta} \partial_\phi \right) (\eta \sin^{-s}\theta),
\nonumber \\
\bar{\eth} \eta &\equiv& - \sin^{-s} \theta \left( \partial_\theta
- \frac{i}{\sin \theta} \partial_\phi \right) (\eta \sin^{s}\theta).
\end{eqnarray}
(The expressions (\ref{eq:splitted}) can be readily obtained
by using the spinor formalism of Ref. \cite{Torres6}.)

We seek separable solutions of Eqs. (\ref{eq:splitted}) of the
form
\begin{eqnarray}
t_{\pm 2} &=& \left[ \frac{j(j+1)}{(j-1)(j+2)}\right]^{1/2}
g_{\pm 2}(r)\,\, _{\pm 2}Y_{jm}(\theta ,\phi), \nonumber \\
t_{\pm 1} &=& \left[ j(j+1) \right]^{1/2}
g_{\pm 1}(r) \,\, _{\pm 1}Y_{jm}(\theta ,\phi),
\label{eq:ansatz} \\
t_{0} &=&  j(j+1) g_{0}(r)\,\, Y_{jm}(\theta ,\phi),\nonumber
\end{eqnarray}
where $j$ is an integer greater than 1 and $_sY_{jm}$ are spin-weighted
spherical harmonics \cite{Newman,Torres5}. The tensor field given by
Eqs. (\ref{eq:ansatz}) is an eigentensor of $J^2$ and $J_3$ with
eigenvalues $j(j+1)$ and $m$, respectively (see, {\it e.g.,}
Ref. \cite{Torres1}). Substituting Eqs. (\ref{eq:ansatz}) into
Eqs. (\ref{eq:splitted}) one obtains the set of ordinary
differential equations
\begin{eqnarray}
\left[ \frac{d^2}{dr^2} +  \frac{2}{r}  \frac{d}{dr} -
\frac{(j-1)(j+2)}{r^2} + k^2 \right]g_{\pm 2}
+ \frac{4(j-1)(j+2)}{r^2} g_{\pm 1} &=& 0, \nonumber \\
\left[ \frac{d^2}{dr^2} +  \frac{2}{r}  \frac{d}{dr} -
\frac{j(j+1) + 4}{r^2} + k^2 \right]g_{\pm 1}
+ \frac{1}{r^2}g_{\pm 2} + \frac{3j(j+1)}{r^2} g_0 &=& 0,
\label{eq:ODE} \\
 \left[ \frac{d^2}{dr^2} +  \frac{2}{r}  \frac{d}{dr} -
\frac{j(j+1) + 6}{r^2} + k^2 \right]g_0
+ \frac{2}{r^2}(g_{- 1} + g_{+ 1}) &=&0. \nonumber
\end{eqnarray}
By combining Eqs. (\ref{eq:ODE}) we find decoupled equations for \cite{Efrain}
\begin{eqnarray*}
g_{+2} - g{-2} &-& 2(j+2) (g_{+1} - g{-1} ) , \\
g_{+2} - g{-2} &+& 2(j-1) (g_{+1} - g{-1} ) , \\
g_{+2} + g{-2} &-& 4(j+2) (g_{+1} + g{-1} ) + 6(j+1)(j+2)g_0, \\
g_{+2} + g{-2} &-& 2 (g_{+1} + g{-1} ) - 2j(j+1)g_0  , \\
g_{+2} + g{-2} &+& 4(j-1)(g_{+1} + g{-1} ) + 6j(j-1)g_0,
\end{eqnarray*}
whose solutions are spherical Bessel functions provided $k\neq 0$.
Thus, from Eqs. (\ref{eq:ansatz}) we get
\begin{eqnarray}
t_{\pm 2}&=& \frac{1}{2}\left[ (j-1)j(j+1)(j+2) \right]^{1/2} \left\{ \,
a \,j_{j+2}(kr) + b\, n_{j+2}(kr) -2 [c \,j_j (kr) +
d\, n_j (kr)] \right. \nonumber \\
&& \left. + e\, j_{j-2}(kr) + f\, n_{j-2}(kr) \pm 2\left[ - A\, j_{j+1}(kr)
- B\, n_{j+1}(kr) \right.  \right.
\nonumber \\
&&
\left. \left.
+ C\, j_{j-1}(kr) + D \,n_{j-1}(kr) \,\right] \right\}\,_{\pm 2} Y_{jm},
\nonumber \\
t_{\pm 1}&=& \frac{1}{2}\left[ j(j+1) \right]^{1/2} \left\{ \, - (j+2)
\left[ a \,j_{j+2}(kr) + b\, n_{j+2}(kr) \right]  +  c \,j_j (kr)
+ d\, n_j (kr) \right.
\nonumber \\
&&  + (j-1) \left[ e\, j_{j-2}(kr) + f\, n_{j-2}(kr) \right]
\pm  (j+2)\left[ A\, j_{j+1}(kr) + B\, n_{j+1}(kr)\right]
\label{eq:solutions} \\
&&  \left.
\pm (j-1)\left[ C\, j_{j-1}(kr) + D \,n_{j-1}(kr)
\,\right] \right\}\, _{\pm 1} Y_{jm},
\nonumber \\
t_0 &=& \left\{ \frac{(j+1)(j+2)}{2} \left[ a\,j_{j+2}(kr)
+ b\, n_{j+2}(kr) \right] + \frac{j(j+1)}{3}\left[ c \,j_j (kr)
+ d\, n_j (kr) \right] \right.
\nonumber \\
&& + \left. \frac{j(j-1)}{2} \left[ e\, j_{j-2}(kr)
+ f\, n_{j-2}(kr)  \right] \right\} \,
Y_{jm}, \nonumber
\end{eqnarray}
where $a,b,c,d,e,f,A,B,C,$ and $D$ are arbitrary constants.

The cases $j=1$ and $j=0$ must be treated separately since $_s Y_{jm}=0$ for
$|s|> j$. We find that, also in these cases, the separable solutions
of Eqs. (\ref{eq:splitted})
are given by Eqs. (\ref{eq:solutions}).

As in the case of the vector Helmholtz equation, the fact that the radial
equations can be decoupled is related with the existence of an operator
that commutes with $J^2$, $J_3$ and $\nabla^2 + k^2$ \cite{Torres2}.
Such an operator can be chosen as
\begin{equation}
K \equiv L_k S_k + 2 ,
\end{equation}
where $L_k$ and $S_k$ are the operators corresponding to the cartesian components
of the orbital and spin angular momentum, respectively, and the summation
convention applies.

For a symmetric, traceless tensor field $t$, the spin-weighted components
of $Kt$ are given by
\begin{eqnarray}
(Kt)_{+2} &=& 2 \eth t_{+1}, \nonumber \\
(Kt)_{+1} &=& -\frac{1}{2} \bar{\eth} t_{+2} - 3 t_{+1} +
\frac{3}{2} \eth t_{0}, \nonumber \\
(Kt)_{0} &=& - \bar{\eth} t_{+1} - 4 t_{0} +
 \eth t_{-1},
\label{eq:solutionsK} \\
(Kt)_{-1} &=& \frac{1}{2} \eth t_{-2} - 3 t_{-1} -
\frac{3}{2} \bar{\eth} t_{0}, \nonumber \\
(Kt)_{-2} &=& -2 \bar{\eth} t_{-1}. \nonumber
\end{eqnarray}
Using Eqs. (\ref{eq:ansatz}) (which give the eigentensors of $J^2$ and $J_3$) and
(\ref{eq:solutionsK}) one can readily find the common eigentensors of $J^2, J_3$
and $K$; for these tensor fields only one of the five radial functions
$g_s (r)$  is independent and, therefore, Eqs. (\ref{eq:ODE}) reduce to a single
equation ({\it cf.} Ref. \cite{Torres2}).

Since $J^2 = L^2 + 2L_k S_k + S^2$ and, for spin-2 fields, $S^2t = 6t$, it follows
that
\begin{equation}
K = \frac{1}{2} (J^2 - L^2 - 2),
\label{eq:newK}
\end{equation}
which shows that the eigentensors of $J^2, S^2,$ and $K$ are also eigentensors
of $L^2$ and using Eqs. (\ref{eq:solutionsK}-\ref{eq:newK}) it is easy to see
that the separable solution (\ref{eq:solutions}) is a superposition of five
eigentensors of $L^2$ with eigenvalues $l(l+1)$, where $l$ coincides with the index
of the spherical Bessel functions appearing in Eqs. (\ref{eq:solutions}).
Assuming that under the parity transformation, ${\mbox{\bf r}} \rightarrow - {\mbox{\bf r}}$,
$\mbox{\bf e}_r$ and $\mbox{\bf e}_\phi$ are left unchanged and $\mbox{\bf e}_\theta$
changes sign and taking into account that $_sY_{jm}$ is transformed
into $(-1)^j \,\,_{-s}Y_{jm}$, one finds that the five eigentensors of
$L^2$ contained in Eqs. (\ref{eq:solutions}) are also eigentensors of
the parity operator with eigenvalue $(-1)^l$.

The divergence of a second-rank, symmetric, traceless tensor field, $t$, is
the vector field div $t$ whose {\it cartesian} components are given by $(\mbox{div}\, t)_i
= \partial_j t_{ij}$, where $\partial_j \equiv \partial /\partial x_j $.
The components of the divergence of $t$ with respect to the basis
$\{ {\mbox{\bf e}}_\theta , {\mbox{\bf e}}_\phi ,{\mbox{\bf e}}_r \}$
are determined by [see, {\it e.g.,} Ref. \cite{Torres6}, Eq. (44)]
\begin{equation}
(\mbox{div}\,\, t)_s = \frac{1}{\sqrt{2}} \left\{
-\frac{1}{r} \bar{\eth} t_{s+1} - \frac{2}{r^3}\partial_r(r^3 t_s)
+ \frac{1}{r} \eth t_{s-1}
\right\},
\label{eq:divergence}
\end{equation}
with the spin-weighted components of div $t$ defined as in
Eq. (\ref{eq:efes}). Substituting Eqs. (\ref{eq:solutions})
into Eqs. (\ref{eq:divergence}) and using the recurrence relations
for the spin-weighted spherical harmonics and for the spherical
Bessel functions, one finds that the separable solution of the
Helmholtz equation given by Eqs. (\ref{eq:solutions}) has vanishing
divergence if and only if
\begin{eqnarray}
a &=& \frac{j(2j -1)c}{3(j+2)(2j + 1)}, \quad  b = \frac{j(2j -1)d}{3(j+2)(2j + 1)},
\quad  e = \frac{(j+1)(2j +3)c}{3(j-1)(2j + 1)},
\label{eq:constants}
\\
f &=& \frac{(j+1)(2j +3)d}{3(j-1)(2j + 1)}, \quad A = \frac{j-1}{j+2}C, \quad \quad  \quad \quad
B = \frac{j-1}{j+2}D.  \nonumber
\end{eqnarray}
Substituting Eqs. (\ref{eq:constants}) into Eqs. (\ref{eq:solutions})
and making use of the recurrence relations for the Bessel functions
one gets
\begin{eqnarray}
t_{+2} &=& - \frac{ik}{r^2} \partial_r r^2 \eth \eth \psi_1
+ \frac{1}{2} \left( \frac{1}{r^2}\partial^2 _r r^2 - k^2 \right) \,
\eth \eth \psi_2,   \nonumber \\
t_{+1} &=&  \frac{ik}{2r} \bar{\eth}  \eth \eth \psi_1
- \frac{1}{2r^2} \partial_r r  \bar{\eth}  \eth \eth \psi_2,
\nonumber \\
t_0 &=& \frac{1}{2r^2} \bar{\eth}  \bar{\eth}  \eth \eth \psi_2,
\label{eq:newsolutionsfort} \\
t_{-1} &=&  \frac{ik}{2r} \eth  \bar{\eth} \bar{\eth} \psi_1
+ \frac{1}{2r^2} \partial_r r  \eth  \bar{\eth} \bar{\eth} \psi_2,
\nonumber \\
t_{-2} &=&  \frac{ik}{r^2} \partial_r r^2 \bar{\eth} \bar{\eth} \psi_1
+ \frac{1}{2} \left( \frac{1}{r^2}\partial^2 _r r^2 - k^2 \right) \,
\bar{\eth} \bar{\eth} \psi_2,   \nonumber
\end{eqnarray}
where
\begin{eqnarray}
\psi_1 &\equiv & \frac{i(2j+1)}{k^2(j+2)} \left[ C \, j_j (kr) + D\,n_j (kr) \right]
\,Y_{jm},
\label{eq:potentials0}
\\
\psi_2 &\equiv & \frac{(2j-1)(2j+3)}{3k^2(j-1)(j+2)} \left[ c \, j_j (kr) + d\,n_j (kr) \right]
\,Y_{jm}. \nonumber
\end{eqnarray}
Clearly, the functions $\psi_1$ and $\psi_2$ are separable solutions of the scalar
Helmholtz equation. It may be noticed that Eqs. (\ref{eq:newsolutionsfort})
contain  no reference to the value of $j$. On the other hand, for $j=1,0$, from
Eqs. (\ref{eq:solutions}) and (\ref{eq:divergence}) one finds that if
$k \neq 0$ and div $t=0$ then, necessarily, $t=0$.

Thus, by virtue of the completeness of the spin-weighted spherical
harmonics, any divergenceless solution of the spin-2 Helmholtz
equation (\ref{eq:helmholtz}) can be expressed as a superposition
of separable solutions of the form (\ref{eq:newsolutionsfort}) where,
now, $\psi_1$ and $\psi_2$ are two solutions of the scalar Helmholtz
equation that are superpositions of solutions of the form
(\ref{eq:potentials0}). In view of Eq. (\ref{eq:identity}), if
$\psi_1$ and $\psi_2$ are real then $t$ is real. (The factor $i$
was included in Eqs. (\ref{eq:newsolutionsfort}) in order to produce
this relation.)

The components (\ref{eq:newsolutionsfort}) can be written in terms
of certain tensor operators $U_{ij}, V_{ij}$ \cite{Campbell},
whose {\it cartesian} components are defined by
\begin{equation}
U_{ij} \equiv L_i X_j \psi + L_j X_i \psi, \quad \quad
V_{ij} \equiv \epsilon_{imn} \partial_m U_{nj} (\psi),
\label{eq:UandV}
\end{equation}
where now
\begin{equation}
{\mbox{\bf L}} \equiv {\mbox{\bf r}} \times \nabla , \quad
\quad {\mbox{\bf X}} \equiv \nabla \times {\mbox{\bf L}}
- \nabla .
\end{equation}
It is easy to see that for any well-behaved function $\psi$,
$U_{ij}(\psi)$ and  $V_{ij}(\psi)$ are symmetric, traceless, divergenceless
tensor fields. By computing the spin-weighted components of
$U_{ij}(\psi)$ and  $V_{ij}(\psi)$ with respect to the basis
$\{ {\mbox{\bf e}}_\theta , {\mbox{\bf e}}_\phi ,{\mbox{\bf e}}_r \}$
\cite{Torres6} one finds that the expressions (\ref{eq:newsolutionsfort})
are equivalent to
\begin{equation}
t_{ij} = k U_{ij}(\psi_1)  +  V_{ij}(\psi_2).
\label{eq:tensor}
\end{equation}

Equations (\ref{eq:UandV}) imply that
\begin{equation}
\varepsilon_{imn}\partial_m V_{nj} (\psi) = - U_{ij} (\nabla^2 \psi ),
\label{eq:other-identity}
\end{equation}
and therefore the tensor field (\ref{eq:tensor}) satisfies
\begin{equation}
({\mbox{curl}}\,\, t)_{ij} \equiv  \varepsilon_{imn}\partial_m t_{nj} =
k^2 U_{ij} (\psi_2) + k V_{ij} (\psi_1).
\label{eq:curl}
\end{equation}
According to Eqs. (\ref{eq:newsolutionsfort}), the scalar potentials
generating a divergenceless solution of the spin-2 Helmholtz equation
are determined by
\begin{equation}
\bar{\eth} \bar{\eth} \eth \eth \psi_2 = 2r^2 t_0 ,
\label{eq:fourderivatives}
\end{equation}
and, by comparing Eqs. (\ref{eq:tensor}) and (\ref{eq:curl}),
\begin{equation}
\bar{\eth} \bar{\eth} \eth \eth \psi_1 = \frac{2r^2}{k} ({\mbox{curl}}\,\,t )_0.
\label{eq:curlagain}
\end{equation}

The uselfulness of Eqs. (\ref{eq:fourderivatives}) and
(\ref{eq:curlagain}) can be illustrated by obtaining the
expansion of a circularly polarized plane wave in spherical waves.
The cartesian components of a spin-2 field corresponding
to a circularly polarized plane wave with helicity
$\pm$ propagating in the $z$-direction are given by
\begin{equation}
(t_{ij})= A \left(
\begin{array}{ccc}
1 & \pm i & 0  \\
\pm i & -1 & 0 \\
0  & 0 & 0
\end{array}
\right)\,e^{ikz},
\label{eq:matrix}
\end{equation}
where $A$ is a constant. From Eqs. (\ref{eq:components}) and
(\ref{eq:matrix}) one finds that $2r^2t_0= x_i x_j t_{ij}
=  A(r\sin \theta \,e^{\pm i\phi})^2 e^{ikz}$; therefore, by
using the expansion of $e^{ikz}$ in terms of spherical harmonics
and the recurrence relations for the associated Legendre functions
(see, {\it e.g.,} Ref. \cite{Arfken}), Eq. (\ref{eq:fourderivatives})
yields
\begin{equation}
 \bar{\eth} \bar{\eth} \eth \eth \psi_2 = -\frac{1}{k^2}
\sum \left[ 4\pi (2j + 1)(j -1)j(j+1)(j+2) \right]^{1/2}
\,i^j \,j_j (kr) \,Y_{j,\pm 2},
\end{equation}
hence, we can choose
\begin{equation}
\psi_2 = -\frac{1}{k^2} \sum_{j=2} ^{\infty} \left[
\frac{4\pi (2j + 1)}{(j -1)j(j+1)(j+2)} \right]^{1/2}
\,i^j \,j_j (kr) \,Y_{j,\pm 2}.
\label{eq:psi2}
\end{equation}
Since the tensor field (\ref{eq:matrix}) is such that that
curl \,$t= \pm kt$, from Eqs. (\ref{eq:fourderivatives}) and
(\ref{eq:curlagain}) we see that
\begin{equation}
\psi_1 = \pm \psi_2.
\label{eq:psi1andpsi2}
\end{equation}
By substituting Eqs. (\ref{eq:psi2}) and  (\ref{eq:psi1andpsi2})
into Eqs. (\ref{eq:newsolutionsfort}) or (\ref{eq:tensor}) one
gets the desired expansion ({\it cf.} Ref. \cite{Jackson}).

In the case where $k=0$, the Helmholtz equation reduces to the
Laplace equation and by assuming a separable solution of the
form (\ref{eq:ansatz}) we obtain [{\it cf.} Eqs. (\ref{eq:solutions})]
\begin{eqnarray}
t_{\pm 2}&=& \frac{1}{2}\left[ (j-1)j(j+1)(j+2) \right]^{1/2} \left\{ \,
a \,r^{j+2} + b\, r^{-j -3} -2 [c \,r^j + d\, r^{-j -1}] \right. \nonumber \\
&& \left. + e\,r^{j-2} + f\, r^{-j +1} \pm 2\left[ - A\, r^{j+1} - B\, r^{-j -2}
+ C\, r^{j -1} + D \,r^{-j} \,\right] \right\}\, _{\pm 2} Y_{jm}, \nonumber \\
t_{\pm 1}&=& \frac{1}{2}\left[ j(j+1) \right]^{1/2} \left\{ \, - (j+2) \left[
a \,r^{j+2} + b\, r^{-j-3} \right]  +  c \,r^j + d\, r^{-j-1} \right. \nonumber \\
&&  + (j-1) \left[ e\, r^{j-2} + f\, r^{-j+1} \right] \pm  (j+2)\left[
 A\, r^{j+1} + B\, r^{-j-2}\right]
\label{eq:solutionsfork0} \\
&& \left.
\pm (j-1)\left[ C\, r^{j-1} + D \,r^{-j} \,\right] \right\}\, _{\pm 1} Y_{jm}, \nonumber \\
t_0 &=& \left\{ \frac{(j+1)(j+2)}{2} \left[ a\,r^{j+2} + b\, r^{-j-3} \right]
+ \frac{j(j+1)}{3}\left[ c \,r^j  + d\, r^{-j-1} \right] \right. \nonumber \\
&& + \left. \frac{j(j-1)}{2} \left[ e\, r^{j-2} + f\, r^{-j-1}  \right] \right\} \,
Y_{jm}, \nonumber
\end{eqnarray}
for $j=0,1,2,\ldots$

Substituting expressions (\ref{eq:solutionsfork0}) into Eqs.
(\ref{eq:divergence}) one finds that the solution  of the Laplace
equation given by Eqs. (\ref{eq:solutionsfork0}) has vanishing
divergence if and only if
\begin{equation}
a=c=d=f=A=D=0.
\label{eq:nullconstants}
\end{equation}
When these relations are inserted into Eqs. (\ref{eq:solutionsfork0}),
for $j > 1$, they can be written as [{\it cf.} Eqs.
(\ref{eq:newsolutionsfort})]
\begin{eqnarray}
t_{+2} &=& - \frac{i}{r^2} \partial_r r^2 \eth \eth \psi_1
+  \frac{1}{2r^2}\partial^2 _r r^2  \,
\eth \eth \psi_2,   \nonumber \\
t_{+1} &=&  \frac{i}{2r} \bar{\eth}  \eth \eth \psi_1
- \frac{1}{2r^2} \partial_r r  \bar{\eth}  \eth \eth \psi_2,
\nonumber \\
t_0 &=& \frac{1}{2r^2} \bar{\eth}  \bar{\eth}  \eth \eth \psi_2,
\label{eq:newsolutionsfortk0} \\
t_{-1} &=&  \frac{i}{2r} \eth  \bar{\eth} \bar{\eth} \psi_1
+ \frac{1}{2r^2} \partial_r r  \eth  \bar{\eth} \bar{\eth} \psi_2,
\nonumber \\
t_{-2} &=&  \frac{i}{r^2} \partial_r r^2 \bar{\eth} \bar{\eth} \psi_1
+ \frac{1}{2r^2} \partial^2 _r r^2 \,
\bar{\eth} \bar{\eth} \psi_2,   \nonumber
\end{eqnarray}
where
\begin{eqnarray}
\psi_1 &= & \frac{i}{(j-1)(j+2)} \left[ (j-1) C \, r^j +
(j+2)B\,r^{-j-1} \right]
\,Y_{jm},
\label{eq:potentialsk0}
\\
\psi_2 &= & \frac{1}{(j-1)(j+2)} \left[ \frac{j-1}{j+1}e \, r^j +
\frac{j+2}{j}b\,r^{-j-1} \right]
\,Y_{jm}. \nonumber
\end{eqnarray}
In this case, $\psi_1$ and $\psi_2$ are separable solutions of the scalar
Laplace equation.

From Eqs. (\ref{eq:solutionsfork0}-\ref{eq:nullconstants}) we see
that the divergenceless solutions of the Laplace equation
with $j=1$ are given by
\begin{equation}
j=1 \quad : \left\{
\begin{array}{ccl}
t_{\pm 2}&=& 0,  \\
t_{\pm 1}&=& \frac{3}{\sqrt{2}} \left( \pm B r^{-3} - b r^{-4} \right)
\, _{\pm 1}Y_{1m}, \\
t_0 &=& 3br^{-4} Y_{1m},
\end{array}
\right.
\end{equation}
which can be written in the form (\ref{eq:newsolutionsfortk0}) with
\begin{eqnarray}
\psi_1 &= & - \frac{iB}{r^2} (Y_{1m} {\mbox{ln}}\, r + 3 h_m ), \nonumber \\
\psi_2 &= & - \frac{b}{r^2} (Y_{1m} {\mbox{ln}} \,r + 3 h_m ),
\label{eq:potentials}
\end{eqnarray}
where $h_m (\theta,\phi)$ is a solution of
\begin{equation}
\bar{\eth} \eth h_m + 2 h_m = Y_{1m} , \quad \quad (m = \pm 1,0) .
\label{eq:hm}
\end{equation}
Owing to Eq. (\ref{eq:hm}), the scalar potentials (\ref{eq:potentials})
satisfy the Laplace equation. Finally, in the case where $j=0$. Eqs.
(\ref{eq:solutionsfork0}-\ref{eq:nullconstants}) yield
\begin{equation}
j=0 \quad : \left\{
\begin{array}{ccc}
t_{\pm 2}&=& 0,  \\
t_{\pm 1}&=& 0, \\
t_0 &=& br^{-3} Y_{00}.
\end{array}
\right.
\end{equation}
This solution is of the form (\ref{eq:newsolutionsfortk0}) with
\begin{equation}
\psi_1 = 0, \quad \quad \psi_2 = \frac{b}{r} Y_{00} \,{\mbox{ln}}\,
\,(r \csc \theta ),
\label{eq:finalpotentials}
\end{equation}
which are solutions of the Laplace equation. It may be noticed
that the scalar potentials (\ref{eq:potentials}) and (\ref{eq:finalpotentials})
diverge at $\theta = 0,\pi,$ and are not separable.

Thus, any divergenceless solution of the spin-2 Laplace equation
can be expressed in the form  (\ref{eq:newsolutionsfortk0}), where
$\psi_1$ and $\psi_2$ are solutions of the scalar
Laplace equation. Equations (\ref{eq:newsolutionsfortk0}) are equivalent
to
\begin{equation}
t_{ij} = U_{ij} (\psi_1) + V_{ij} (\psi_2).
\label{eq:tensoragain}
\end{equation}

\section{Separation of variables in cylindrical coordinates}

Taking now the basis $\{ {\mbox{\bf e}}_1 , {\mbox{\bf e}}_2 ,{\mbox{\bf e}}_3 \}$
as the orthonormal basis  $\{ {\mbox{\bf e}}_\rho ,
{\mbox{\bf e}}_\phi ,{\mbox{\bf e}}_z \}$
induced by the circular cylindrical coordinates, the Helmholtz
equation for a symmetric, traceless, second-rank tensor field is
equivalent to the set of uncoupled equations
\begin{equation}
\partial^2 _z t_s + \bar{\eth}\eth t_s + k^2 t_s = 0, \quad \quad
(s= \pm 2,\pm1,0),
\label{eq:hcylindrical}
\end{equation}
where acting on a quantity $\eta$ with spin-weight $s$ \cite{Torres3},
\begin{eqnarray}
\eth \eta &\equiv & - \rho^s \left( \partial_\rho + \frac{i}{\rho}
\partial_\phi \right) (\rho^{-s}\eta), \nonumber \\
\bar{\eth} \eta &\equiv & - \rho^{-s} \left( \partial_\rho - \frac{i}{\rho}
\partial_\phi \right) (\rho^{s}\eta).
\end{eqnarray}
We seek solutions of Eqs. (\ref{eq:hcylindrical}) of the form
\begin{equation}
t_s = g_s (z) \, _sF_{\alpha m}(\rho,\phi), \quad \quad (s= \pm 2,\pm1,0),
\label{eq:ansatzcylindrical}
\end{equation}
where the $_sF_{\alpha m}$ are spin-weighted cylindrical harmonics
 \cite{Torres3}. A tensor field of the form (\ref{eq:ansatzcylindrical})
is eigentensor of $J_3$ and of the square of the linear momentum perpendicular
to the $z$-axis, $P_1 ^2 + P_2 ^2$, with eigenvalues $m$ and $\alpha^2$,
respectively \cite{Torres3}. Substituting Eqs. (\ref{eq:ansatzcylindrical})
into Eqs. (\ref{eq:hcylindrical}) one finds
\begin{equation}
\left(  \frac{d^2}{dz^2} - \gamma^2 \right)\,g_s = 0, \quad \quad
(s= \pm 2,\pm1,0),
\end{equation}
where
\begin{equation}
\gamma^2 \equiv \alpha^2 - k^2,
\end{equation}
therefore, if $\gamma \neq 0$, $g_s (z) = A_s e^{\gamma z} + B_s e^{-\gamma z}$
and if $\gamma = 0$, $g_s (z) = A_s + B_s z$, where $A_s$ and $B_s$ are
arbitrary constants. Thus, assuming that $\alpha$ is different from zero,
Eqs. (\ref{eq:hcylindrical}) admit separable solutions of the form
\begin{equation}
t_s = (A_s e^{\gamma z} + B_s e^{-\gamma z})[C_s ( _sJ_{\alpha m})
+ D_s( _sN_{\alpha m})], \quad \quad (\gamma \neq 0),
\label{eq:solutions-a}
\end{equation}
and
\begin{equation}
t_s = (A_s  + B_s  z )[C_s ( _sJ_{\alpha m})
+ D_s( _sN_{\alpha m})], \quad \quad (\gamma = 0),
\label{eq:solutions-b}
\end{equation}
where $A_s , B_s , C_s$ and $D_s$ are arbitrary constants and \cite{Torres3}
\begin{equation}
_sJ_{\alpha m}(\rho,\phi) \equiv J_{m+s}(\alpha \rho)e^{im\phi},
\quad \quad
_sN_{\alpha m}(\rho,\phi) \equiv N_{m+s}(\alpha \rho)e^{im\phi},
\end{equation}
with $J_\nu$ and $N_\nu$ being Bessel functions. For $\alpha = 0 $, the
functions $_sF_{\alpha m}$ diverge when $\rho$ goes to zero or to infinity,
or they do not vanish when $\rho$ goes to infinity.

The components of the divergence of a symmetric, traceless, second-rank
tensor field with respect to the basis $\{ {\mbox{\bf e}}_\rho ,
{\mbox{\bf e}}_\phi ,{\mbox{\bf e}}_z \}$ are given by [see, {\it e.g.,}
Ref. \cite{Torres6}, Eq. (44)]
\begin{equation}
({\mbox{div}}\,t)_s = \frac{1}{\sqrt{2}} \left\{
- \bar{\eth}t_{s+1} - 2 \partial_z t_s + \eth t_{s-1} \right\},
\end{equation}
therefore, the tensor field  (\ref{eq:solutions-a}) has vanishing
divergence if and only if
\begin{equation}
\begin{array}{cc}
\frac{\alpha}{2}(A_0 C_0 + A_{\pm2}C_{\pm2}) = \gamma A_{\pm 1}C_{\pm 1}, &
\frac{\alpha}{2}(A_0 D_0 + A_{\pm2}D_{\pm2}) = \gamma A_{\pm 1}D_{\pm 1},\\
\frac{\alpha}{2}(B_0 C_0 + B_{\pm2}C_{\pm2}) = -\gamma B_{\pm 1}C_{\pm 1}, &
\frac{\alpha}{2}(B_0 D_0 + B_{\pm2}D_{\pm2}) = -\gamma B_{\pm 1}D_{\pm 1}, \\
\frac{\alpha}{2}(A_{+1} C_{+1} + A_{-1}C_{-1}) = \gamma A_{0}C_{0}, &
\frac{\alpha}{2}(A_{+1} D_{+1} + A_{-1}D_{-1}) = \gamma A_{0}D_{0}, \\
\frac{\alpha}{2}(B_{+1} C_{+1} + B_{-1}C_{-1}) = -\gamma B_{0}C_{0}, &
\frac{\alpha}{2}(B_{+1} D_{+1} + B_{-1}D_{-1}) = -\gamma B_{0}D_{0}.
\end{array}
\label{eq:relations1}
\end{equation}
Introducing the combinations
\begin{equation}
\begin{array}{cc}
a_1 \equiv \frac{1}{2\alpha}(A_{+1} C_{+1} - A_{-1}C_{-1}), &
a_2 \equiv \frac{1}{2\alpha}(A_{+1} D_{+1} - A_{-1}D_{-1}), \\
b_1 \equiv \frac{1}{2\alpha}(B_{+1} C_{+1} - B_{-1}C_{-1}), &
b_2 \equiv \frac{1}{2\alpha}(B_{+1} D_{+1} - B_{-1}D_{-1}), \\
a_3 \equiv \frac{1}{2\alpha \gamma}(A_{+1} C_{+1} + A_{-1}C_{-1}), &
a_4 \equiv \frac{1}{2\alpha \gamma}(A_{+1} D_{+1} + A_{-1}D_{-1}), \\
b_3 \equiv -\frac{1}{2\alpha \gamma}(B_{+1} C_{+1} + B_{-1}C_{-1}), &
b_4 \equiv -\frac{1}{2\alpha \gamma}(B_{+1} D_{+1} + B_{-1}D_{-1}),
\end{array}
\label{eq:relations2}
\end{equation}
and assuming that the conditions (\ref{eq:relations1}) hold, one finds that
the components (\ref{eq:solutions-a}) can be written as [{\it cf.} Eqs.
(\ref{eq:newsolutionsfort}) and (\ref{eq:newsolutionsfortk0})]
\begin{eqnarray}
t_{+2} &=& - i \partial_z  \eth \eth \psi_1
+  \frac{1}{2}(\partial^2 _z - k^2)  \,
\eth \eth \psi_2,   \nonumber \\
t_{+1} &=&  \frac{i}{2} \bar{\eth}  \eth \eth \psi_1
- \frac{1}{2} \partial_z   \bar{\eth}  \eth \eth \psi_2,
\nonumber \\
t_0 &=& \frac{1}{2} \bar{\eth}  \bar{\eth}  \eth \eth \psi_2,
\label{eq:newsolutionsfortk0-c} \\
t_{-1} &=&  \frac{i}{2} \eth  \bar{\eth} \bar{\eth} \psi_1
+ \frac{1}{2} \partial_z   \eth  \bar{\eth} \bar{\eth} \psi_2,
\nonumber \\
t_{-2} &=&  i \partial_z  \bar{\eth} \bar{\eth} \psi_1
+  \frac{1}{2}(\partial^2 _z - k^2)\,
\bar{\eth} \bar{\eth} \psi_2,   \nonumber
\end{eqnarray}
where
\begin{eqnarray}
\psi_1 &= & \frac{2i}{\alpha^2} \left[ (a_1 e^{\gamma z} +
b_1 e^{-\gamma z})\, _0J_{\alpha m} +
(a_2 e^{\gamma z} + b_2 e^{-\gamma z})\, _0N_{\alpha m} \right],
\label{eq:potentialsk0-c}
\\
\psi_2 &= & \frac{2}{\alpha^2} \left[ (a_3 e^{\gamma z} +
b_3 e^{-\gamma z})\, _0J_{\alpha m} +
(a_4 e^{\gamma z} + b_4 e^{-\gamma z})\, _0N_{\alpha m} \right],
\nonumber
\end{eqnarray}
which are solutions of the scalar Helmholtz equation.

Similarly, one finds that if the field given by Eqs. (\ref{eq:solutions-b})
has vanishing divergence then its components can be written in the form
(\ref{eq:newsolutionsfortk0-c}), where
 \begin{eqnarray}
\psi_1 &= & \frac{2i}{\alpha^2} \left[ (a_1 + b_1 z )\, _0J_{\alpha m} +
(a_2  + b_2 z)\, _0N_{\alpha m} \right],
\label{eq:potentialsk0-c2}
\\
\psi_2 &= & \frac{2}{\alpha^2} \left[ (a_3  + b_3 z)\, _0J_{\alpha m} +
(a_4  + b_4 z)\, _0N_{\alpha m} \right],
\nonumber
\end{eqnarray}
which satisfy the scalar Helmholtz equation, and
\begin{equation}
\begin{array}{cc}
a_1 \equiv \frac{1}{2\alpha}(A_{+1} C_{+1} - A_{-1}C_{-1}), &
a_2 \equiv \frac{1}{2\alpha}(A_{+1} D_{+1} - A_{-1}D_{-1}), \\
b_1 \equiv \frac{1}{2\alpha}(B_{+1} C_{+1} - B_{-1}C_{-1}), &
b_2 \equiv \frac{1}{2\alpha}(B_{+1} D_{+1} - B_{-1}D_{-1}), \\
a_3 \equiv -\frac{1}{2\alpha^2}(A_{+2} C_{+2} + A_{-2}C_{-2}), &
a_4 \equiv -\frac{1}{2\alpha^2}(A_{+2} D_{+2} + A_{-2}D_{-2}), \\
b_3 \equiv \frac{1}{2\alpha }(A_{+1} C_{+1} + A_{-1}C_{-1}), &
b_4 \equiv \frac{1}{2\alpha }(A_{+1} D_{+1} + A_{-1}D_{-1}).
\end{array}
\label{eq:relations3}
\end{equation}
It can be shown that Eqs. (\ref{eq:newsolutionsfortk0-c}) are equivalent
to
\begin{equation}
t_{ij} = W_{ij} (\psi_1) + Z_{ij} (\psi_2),
\label{eq:tensor-c}
\end{equation}
where the tensor operators $W_{ij}$ and $Z_{ij}$ are given in cartesian
coordinates by
\begin{equation}
W_{ij}(\psi) \equiv M_iN_j \,\psi + M_j N_i \,\psi ,\quad \quad  Z_{ij} (\psi)
\equiv \varepsilon_{imn}\partial_m W_{nj} (\psi),
\end{equation}
with
\begin{equation}
{\mbox{\bf M}} \equiv  {\mbox{\bf e}}_z \times \nabla , \quad \quad
{\mbox{\bf N}} \equiv \nabla \times {\mbox{\bf M}} ,
\end{equation}
[{\it cf.} Eqs. (\ref{eq:UandV}-\ref{eq:tensor})]. It is easy to see that for
any well-behaved function $\psi$, $W_{ij}(\psi)$ and $Z_{ij}(\psi)$ are symmetric,
traceless, divergenceless tensor fields, and that
\begin{equation}
\varepsilon_{imn}\partial_m Z_{nj} (\psi) = - W_{ij} (\nabla^2 \psi).
\label{eq:more}
\end{equation}
The solutions (\ref{eq:ansatzcylindrical}) with $\alpha =0$ can also be written
in the form  (\ref{eq:newsolutionsfortk0-c}), in terms of two scalar potentials
$\psi_1$ and $\psi_2$ that satisfy the Helmholtz equation but they are not
separable. Owing to the completeness of the spin-weighted cylindrical harmonics,
any divergenceless solution of the spin-2 Helmholtz equation can be written
in the form (\ref{eq:newsolutionsfortk0-c}), where $\psi_1$ and $\psi_2$ are
solutions of the scalar Helmholtz equation. If $\psi_1$ and $\psi_2$ are real,
then $t_{ij}$ is real.

\section{Application to the linearized Einstein equations}

The Einstein vacuum field equations linearized about the Minkowski metric
can be written in cartesian coordinates in the gauge-invariant form
\begin{eqnarray}
\partial_i E_{ij}&=& 0,\quad \quad \quad  \quad \quad \quad \quad  \partial_i B_{ij} =0,
\label{eq:Maxwell1}
\\
\frac{1}{c} \frac{\partial }{\partial \,t}E_{ij} &=& \varepsilon_{imn}\partial_m B_{nj},
\quad \quad  \frac{1}{c} \frac{\partial }{\partial \,t} B_{ij}= - \varepsilon_{imn}
\partial_m E_{nj},
\label{eq:Maxwell2}
\end{eqnarray}
where $E_{ij}$ and $B_{ij}$ are symmetric traceless tensor fields defined by
\begin{equation}
E_{ij} \equiv K_{0i0j}, \quad \quad B_{ij} \equiv -\frac{1}{2} K_{0imn}\, \varepsilon_{jmn},
\end{equation}
and $K_{\alpha \beta \gamma \delta}$ is the curvature tensor to first order in the metric
perturbation (see {\it e.g.,} Refs. \cite{Campbell, Torres7}). From Eqs. (\ref{eq:Maxwell1},
\ref{eq:Maxwell2}) it follows that the fields  $E_{ij}$ and $B_{ij}$ obey the wave equation;
therefore, assuming that  $E_{ij}$ and $B_{ij}$ have a time dependence of the form
$e^{-i\omega t}$, the fields $E_{ij}$ and $B_{ij}$ satisfy the Helmholtz equation
with $k=\omega /c.$ According to the results of Sect. 2, if $\omega \neq 0$, $E_{ij}$ can
be expressed in the form
\begin{equation}
E_{ij} = k U_{ij} (\psi_1) + V_{ij} (\psi_2),
\label{eq:electric-field}
\end{equation}
[{\it cf.} Eq. (\ref{eq:tensor})] where $\psi_1$ and $\psi_2$ are regular solutions
of the scalar Helmholtz equation. From Eqs. (\ref{eq:Maxwell2}) and (\ref{eq:curl})
we see that the field $B_{ij}$ corresponding to (\ref{eq:electric-field}) is given
by
\begin{equation}
B_{ij} =-i[ k U_{ij} (\psi_2) + V_{ij} (\psi_1)].
\label{eq:magnetic-field}
\end{equation}

In the static case $(\omega = 0)$, $E_{ij}$ and $B_{ij}$ must satisfy the Laplace equation;
hence, $E_{ij}$ can be expressed in the form $E_{ij} = U_{ij} (\psi_1) + V_{ij} (\psi_2)$,
where $\psi_1$ and $\psi_2$ are solutions of the scalar Laplace equation
[Eq. (\ref{eq:tensoragain})]. Equations (\ref{eq:Maxwell2}) and (\ref{eq:other-identity})
give $0= \varepsilon_{imn}\partial_m E_{nj}=V_{ij}(\psi_1)$, which implies that $U_{ij}(\psi_1) = 0$
[see Eqs. (\ref{eq:newsolutionsfortk0}-\ref{eq:finalpotentials})]. Thus
\begin{equation}
E_{ij}=V_{ij}(\psi_2).
\label{eq:E1}
\end{equation}
In a similar manner, it follows that
\begin{equation}
B_{ij}=V_{ij}(\psi_4),
\label{eq:B1}
\end{equation}
where $\psi_4$ is a solution of the scalar Laplace equation. (It may be noticed
that Eqs. (\ref{eq:E1}-\ref{eq:B1}) can be obtained from Eqs.
(\ref{eq:electric-field}-\ref{eq:magnetic-field}) by simply setting $k=0$.)

Alternatively, the components $E_{ij}$ and $B_{ij}$ can be expressed in the form
(\ref{eq:tensor-c}). Assuming again that the time dependence of the fields is
given by a factor  $e^{-i\omega t}$, in the case where $\omega \neq 0$,
\begin{equation}
E_{ij} = k W_{ij} (\psi_1) + Z_{ij} (\psi_2),
\label{eq:electric-field2}
\end{equation}
where $\psi_1$ and $\psi_2$ are solutions of the scalar Helmholtz
equation and the factor $k$ has been introduced for convenience.
Then, Eqs. (\ref{eq:Maxwell2}) and (\ref{eq:more}) imply that
\begin{equation}
B_{ij} =-i[ k W_{ij} (\psi_2) + Z_{ij} (\psi_1)].
\label{eq:magnetic-field2}
\end{equation}
On the other hand, when $\omega = 0$,
\begin{equation}
E_{ij} = Z_{ij} (\psi_1), \quad \quad B_{ij} = Z_{ij} (\psi_2),
\label{eq:last}
\end{equation}
where $\psi_1$ and $\psi_2$ are solutions of the scalar Laplace equation.

In the standard approach, the Einstein vacuum field equations
linearized about the Minkowski metric are written in terms of the metric
perturbations, which are affected by the gauge transformations induced
by the infinitesimal coordinate changes. By contrast, the curvature
perturbations $K_{\alpha \beta \gamma \delta}$, which are equivalent
to the fields $E_{ij}$ and $B_{ij}$, provide a gauge-invariant description
of the gravitational field Eqs. (\ref{eq:electric-field}-\ref{eq:magnetic-field}),
which are adapted to the spherical coordinates, yield a multipole expansion
of the gravitational field. The gravitational potentials appearing
in Eqs. (\ref{eq:electric-field}-\ref{eq:last}) for fields generated
by localized sources can be expressed in terms of the energy-momentum
tensor of the sources as in Refs. \cite{Campbell,Torres4}.

\section{Concluding remarks}

It is known that in flat space-time a massless field of arbitrary
spin can be expressed in terms of two real potentials or of a single
complex scalar potential (see, {\it e.g.,} Ref. \cite{Penrose});
the results presented above and in Refs. \cite{Torres1,Torres3} show
specifically that there exist operators such that when applied to a
solution of the scalar Helmholtz equation which is eigenfunction
of $J^2 (=L^2)$ and $J_3 (=L_3)$, they yield solutions of the spin-1
and spin-2 massless field equations that are eigenfunctions of $J^2$
and $J_3$ with the eigenvalues of the potential and that, similarly,
there exist operators that map a solution of the scalar Helmholtz
equation which is eigenfunction of $P_1 ^2 + P_2 ^2$ and $J_3$ into
solutions of the spin-1 and spin-2 massless field equations that
are eigenfunctions of $P_1 ^2 + P_2 ^2$ and $J_3$ with the same
respective eigenvalues.

 \vfill\break
\bibliographystyle{unsrt}

\end{document}